\documentclass[11pt,a4paper]{article}
\pdfoutput=1
\usepackage{jheppub}

\usepackage{color}
\usepackage{amsmath}
\usepackage{pifont}

\usepackage[utf8]{inputenc}

\usepackage{verbatim}   
\usepackage{subfigure}  

\usepackage{amsfonts}
\usepackage{amssymb}
\usepackage{mathrsfs}
\usepackage{graphicx}
\usepackage{multirow}
 \usepackage{slashed}

\definecolor{mygreen}{RGB}{29,145,47}
\definecolor{mypurple}{RGB}{164,64,214}
\definecolor{myorange}{RGB}{199,146,32}

\newcommand{\cm}{{\, {\rm cm}}}

\newcommand{\MeV}{{\, {\rm MeV}}}
\newcommand{\GeV}{{\, {\rm GeV}}}
\newcommand{\TeV}{{\, {\rm TeV}}}

\def\beq{\begin{equation}}
\def\eeq{\end{equation}}
\def\bea{\begin{eqnarray}}
\def\eea{\end{eqnarray}}
\def\bitem{\begin{itemize}}
\def\eitem{\end{itemize}}
\newcommand{\bec}{\begin{center}}
\newcommand{\eec}{\end{center}}
\newcommand{\ba}{\begin{array}}
\newcommand{\ea}{\end{array}}

\def\bar#1{\overline{#1}}

\def\inv{^{\raise.15ex\hbox{${\scriptscriptstyle -}$}\kern-.05em 1}}
\def\lbar{{\lower.35ex\hbox{$\mathchar'26$}\mkern-10mu\lambda}} 

\let\<=\langle
\let\>=\rangle

\let\+=\uparrow

\begin{document}

\title{Higgs portal dark matter in non-thermal cosmologies} 
\author[a]{Edward Hardy}
\emailAdd{ehardy@liverpool.ac.uk}
\affiliation[a]{Department of Mathematical Sciences, University of Liverpool, Liverpool, L69 7ZL, United Kingdom}

\abstract{
A scalar particle with a relic density set by annihilations through a Higgs portal operator is a simple and minimal possibility for dark matter. However, assuming a thermal cosmological history this model is ruled out over most of parameter space by collider and direct detection constraints. We show that in theories with a non-thermal cosmological history Higgs portal dark matter is viable for a wide range of dark matter masses and values of the portal coupling, evading existing limits. In particular, we focus on the string theory motivated scenario of a period of late time matter domination due to a light modulus with a decay rate that is suppressed by the Planck scale. Dark matter with a mass $\lesssim \GeV$ is possible without additional hidden sector states, and this can have astrophysically relevant self-interactions. We also study the signatures of such models at future direct, indirect, and collider experiments. Searches for invisible Higgs decays at the high luminosity LHC or an $e^+$ $e^-$ collider could cover a significant proportion of the parameter space for low mass dark matter, and future direct detection experiments will play a complementary role.
}

\maketitle



\section{Introduction} \label{sec:intro}

Higgs portal dark matter (DM) \cite{Silveira:1985rk,McDonald:1993ex,Burgess:2000yq,Patt:2006fw} with a relic abundance set by freeze out during a standard, thermal, cosmological history is a minimal and predictive model. In its simplest implementation only one new state is introduced, a scalar $\chi$ that is uncharged under the Standard Model (SM) gauge group. This is stabilised with a $Z_2$ symmetry $\chi \rightarrow -\chi$, and couples to the SM Higgs field $H$ through the renomalisible interaction\footnote{The phenomenology is very similar if the DM is a complex or real scalar, we take it to be real throughout.}
\beq \label{eq:1}
\mathcal{L} \supset -\frac{\lambda}{2} \chi^2 H^{\dagger} H ~.
\eeq
In models with a thermal cosmological history $\chi$ is in thermal equilibrium with the visible sector in the early universe, before freezing out and acting as DM.\footnote{This is assuming the coupling $\lambda$ is not very small. If $\lambda$ is tiny,  $ \sim 10^{-10}$, the DM relic abundance could be generated from freeze in \cite{McDonald:2001vt,Hall:2009bx}.} Vector or fermion DM with a relic abundance set by annihilations through a Higgs portal operator is also possible, although generating a mass for the former requires a hidden sector Higgs field or the Stueckelberg mechanism, and in the latter case the portal operator is non-renormalisible and additional new states must be introduced at an intermediate energy scale \cite{Kim:2006af,Kim:2008pp,LopezHonorez:2012kv} .

While their simplicity is appealing, minimal Higgs portal models are severely constrained by collider limits on the invisible branching fraction of the Higgs and direct detection searches, most recently at Xenon1T \cite{Aprile:2017iyp} and PandaX2 \cite{Cui:2017nnn}. As we discuss in Section~\ref{sec:hp}, after imposing that the relic density matches the observed value, scalar or vector Higgs portal DM is only possible for a limited range of DM masses \cite{Escudero:2016gzx}.  
 Only fermion DM with pseudoscalar couplings to the Higgs remains viable over significant parts of parameter space, since this has an extremely suppressed signal at direct detection experiments.

However, there is no strong reason to think that the cosmological history of the universe before big bang nucleosynthesis (BBN) was thermal, either observationally or from a model building perspective. Instead, in typical string theory models there are many scalar moduli fields, the mass of which is typically set by the gravitino mass  \cite{Coughlan:1983ci,Goncharov:1984qm,Ellis:1986zt,deCarlos:1993wie,Banks:1993en}. If this is fairly low, motivated for example by the success of gauge unification in supersymmetric extensions of the SM, the moduli will dominate the energy density of the universe at early times leading to an extended period of matter domination. The moduli eventually decay through Planck suppressed operators, and a radiation dominated universe re-emerges before BBN, consistent with measurements of the abundances of light elements.

It is well known that altering the cosmological history of the universe increases the parameter space over which the required  relic density can be obtained in a wide variety of DM models \cite{Kamionkowski:1990ni,Giudice:2000ex}. For example, this has previously been considered for weakly interacting massive particles (WIMPs) \cite{Moroi:1999zb,Gelmini:2006pw,Gelmini:2006pq,Acharya:2009zt}, more general DM hidden sector candidates \cite{Kane:2015qea,Acharya:2017kfi,Drees:2017iod}, hidden sector glueballs \cite{Acharya:2017szw}, and axions \cite{Kim:1992eu,Lyth:1993zw,Kawasaki:1995vt,Grin:2007yg,Visinelli:2009kt,Acharya:2010zx}. In this paper we show that allowing a non-thermal cosmological history also significantly increases the parameter space for minimal Higgs portal DM models. By adjusting the reheating temperature after the decay of the lightest modulus, and the branching fraction of the lightest modulus into DM, all points in the plane of DM mass ($m_{\rm DM}$) against Higgs portal interaction strength with $m_{\rm DM} \gtrsim \GeV$ can lead to the observed relic abundance, and there is a significant region of viable parameter space with $100~\MeV \lesssim  m_{\rm DM} \lesssim \GeV$. 

Although an increase in the range of allowed DM masses and values of the portal coupling is not surprising, it is interesting that low mass DM is possible in simple models. Such DM candidates can have self interactions that are relevant for dynamics on galactic scales (for $m_{\rm DM} \gtrsim \GeV$ perturbativity bounds mean that significant self interactions are possible only if additional light mediators are introduced \cite{Tulin:2013teo}). Astrophysically relevant self interactions could resolve conflicts between simulations of cold DM and observations of small scale structure, although these may also be due to incomplete inclusion of baryonic physics in simulations. In contrast, in almost all models with a thermal cosmological history indirect detection searches already rules out DM with a mass $\lesssim \GeV$ and a relic density set by annihilations to the visible sector \cite{Chu:2016pew}.\footnote{The exception is a Majorana fermion or scalar DM candidate interacting with the visible sector through a new $Z'$ gauge boson, with mass $\sim \GeV$. Despite strong constraints from measurements of $\Delta N_{\rm eff}$ by Planck and from fixed target experiments \cite{Batell:2009di} such models remain viable in a small part of parameter space.} Instead, models with a thermal cosmology and DM in this mass range typically require additional new states for the DM to annihilate into. As well as involving more complex model building \cite{Feng:2008mu} (for example, to avoid the extra particles over-closing the universe themselves or violating constraints on the effective number of relativistic degrees of freedom in the early universe), this removes the link between the DM annihilation rate to the visible sector and its relic abundance. As a result, the motivation for a coupling between the visible and DM sectors, and therefore signals that could be probed by future experiments, to be present at all is diminished. 

Higgs portal DM can also lead to a diverse range of observational signatures in direct, indirect, and collider experiments. In return for making the cosmological history of the universe more complicated, albeit in a way that is motivated from UV considerations, the models that we consider can lead to signals in parts of parameter space where they would otherwise not be expected (unless a more complex hidden sector is introduced). For example, we will show that there are many non-thermal models with $m_{\rm DM} < m_h / 2$ that could be explored by searches for invisible Higgs decays at the high luminosity LHC. 

Turning to the structure of this paper, in the next section we study the status of Higgs portal DM models assuming a thermal cosmology. Following this, in Section~\ref{sec:nontherm} we discuss the possibility of a period of late time matter domination, and the motivation for this from string theory. In Section~\ref{sec:hpnon} we study the viable parameter space for Higgs portal DM in such theories, and in Section~\ref{sec:future} we consider the potential future experimental signatures. Finally, in Section~\ref{sec:con} we summarise our the results.

\section{Higgs portal DM with a thermal cosmological history}\label{sec:hp}

We focus on the case of a real scalar DM candidate, and require that this makes up the full measured DM relic abundance. For comparison with theories with late time matter domination, we first analyse this model assuming a thermal cosmological history. Rather than attempting to fully explore the uncertainties on all of the relevant experimental searches, it is sufficient for our present purposes to summarise the broad features of the constraints. Additional recent discussion and reviews may be found in, for example, \cite{Assamagan:2016azc,Athron:2017kgt}.

\begin{figure}
\begin{center}
 \includegraphics[width=0.75\textwidth]{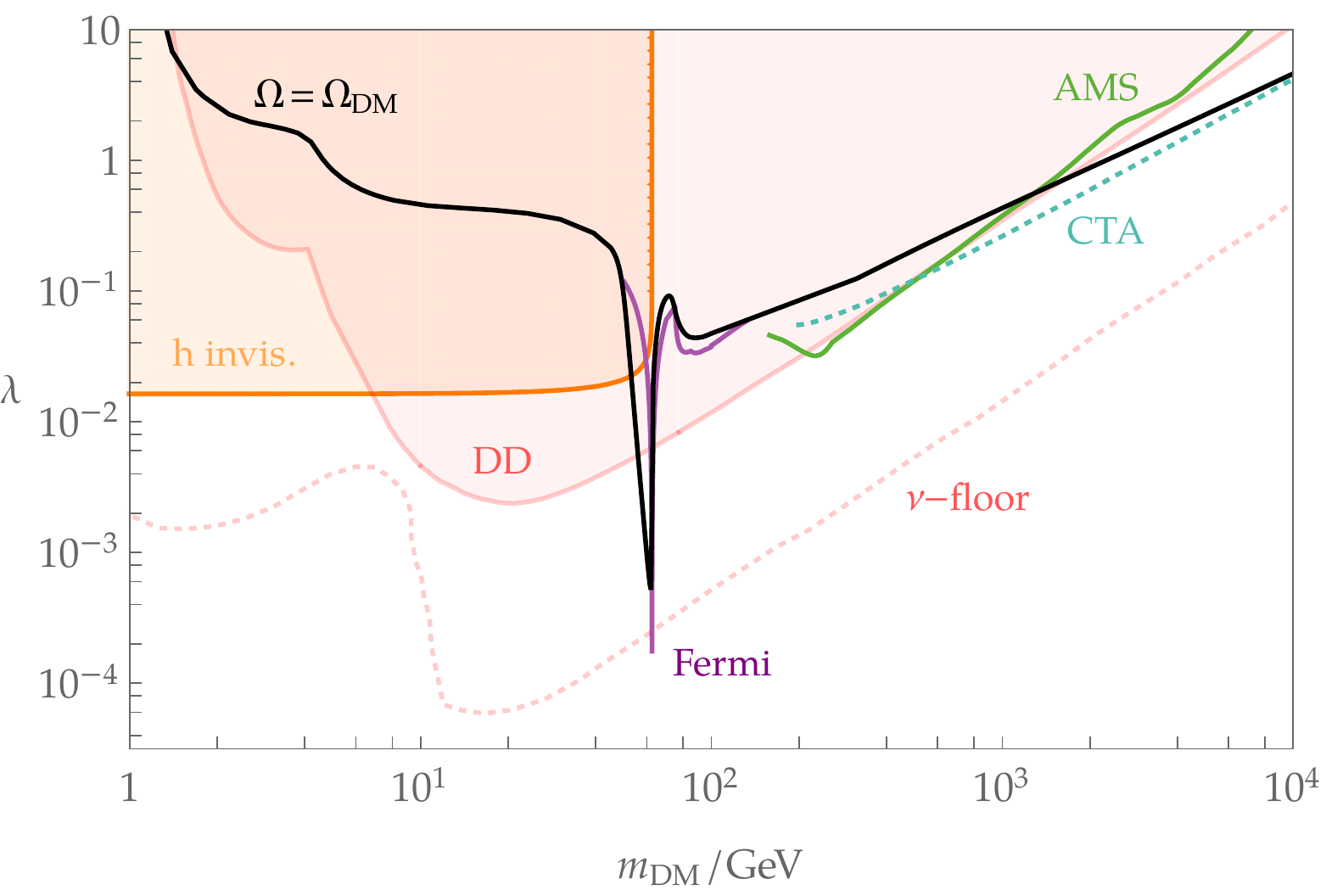}
\caption{The constraints on scalar Higgs portal DM models assuming a thermal cosmological history. Requiring that the relic abundance matches the measured value fixes the value of portal coupling as a function of the DM mass to the contour labeled $\Omega = \Omega_{\rm DM}$. Direct detection experiments (DD) lead to strong constraints, and LHC searches for invisible Higgs decays are also relevant (h invis.). Indirect detection searches by Fermi and AMS are potentially constraining, although the latter is highly sensitive to astrophysical uncertainties (these are plotted only in parts of parameter space in which they are competitive with direct detection searches). The reach of the future CTA experiment is shown, and this could probe the high DM mass region. The neutrino floor ($\nu$-floor), below which direct detection searches become extremely challenging, is also shown.}
\label{fig:thermal} 
\end{center} 
\end{figure}

In this simple model there are only two free parameters that affect the DM relic abundance: its mass $m_{\rm DM}$ and the value of the portal coupling $\lambda$, defined in Eq.~\eqref{eq:1}. As a result, the precise measurement of the DM relic abundance by Planck \cite{Ade:2015xua} fixes the required value of $\lambda$ for a given DM mass. The computation of the relic abundance in a thermal cosmology is standard \cite{Cline:2013gha,Escudero:2016gzx,Binder:2017rgn}, and we obtain our results by numerically integrating the Boltzmann equations for the DM number density. The possibility of resonant annihilation through the SM Higgs means that the thermal average of the annihilation rate must be carried out numerically, rather than using an analytic approximation. To find the annihilation cross section, we follow \cite{Cline:2013gha} in taking the 1-loop corrected cross section at large DM masses, and using results computed for the SM Higgs width at smaller masses $m_{\rm DM} < 300~\GeV$ \cite{Dittmaier:2011ti} (with the addition of the extra channel into the SM Higgs when this is kinematically allowed). The tabulated values of the Higgs width include the leading QCD corrections and the effects of 3 and 4 body final states, which can change the cross section significantly \cite{Cline:2012hg}.\footnote{There might be a relatively minor inaccuracy in our calculation of the relic abundance for DM masses in the range where resonant annihilation through the SM Higgs occurs, if this leads to kinetic decoupling at around the same time as chemical decoupling \cite{Duch:2017nbe}. Given that the viability of DM extremely close to resonance is not our main focus, we neglect this possible effect.} 

The contour of $\lambda$ against $m_{\rm DM}$ that leads to the experimentally measured DM relic abundance is plotted in Figure~\ref{fig:thermal} (this closely matches existing results in the literature). There is a region around $m_{\rm DM} \simeq m_h/2$ in which resonant annihilation permits relatively small values of the portal coupling. However, over the remainder of parameter space couplings $\lambda \gtrsim 0.1$ are required. For DM with a large mass, annihilation is dominantly to $W$-bosons, with significant branching fractions also to $Z$-bosons and the SM Higgs. At times when the DM is not highly relativistic,  annihilation is primarily to b-quarks if $m_{\rm W} > m_{\rm DM} > m_b$, where $m_{\rm W}$ and $m_b$ are the masses of the W-boson and b-quark respectively.

\subsection{Direct, indirect, and collider constraints} \label{sec:searches}

The simple nature of the Higgs portal model also means that, once the portal coupling and DM mass are chosen, the observational signals in direct, indirect, and collider searches are fixed. Uncertainties on the constraints obtained arise from, for example, our limited knowledge of the DM density close to the galactic centre and in the vicinity of the Earth, and the quark content of nucleons.

Over the majority of parameter space, the dominant constraint on scalar Higgs portal models comes from direct detection experiments. The cross section for spin-independent DM scattering off nucleons, labeled $N$, in such models can be parameterised as 
\beq \label{eq:sig}
\sigma = \lambda^2 \frac{f_N^2}{4\pi} \frac{\mu^2 m_N^2}{m_{\rm DM}^2 m_h^4} ~,
\eeq
where $\mu = m_N m_{\rm DM}/\left( m_N + m_{\rm DM} \right)$, with $m_N$ the nucleon mass. At leading order in a chiral expansion 
\begin{equation}
f_N = \sum_q \frac{m_q}{m_N} \left< N \left| \bar{q} q \right| N \right> ~,
\end{equation}
where $q$ denotes the SM quarks \cite{Andreas:2008xy,Cline:2013gha,Beniwal:2015sdl}. Recently the scattering cross section has been computed to higher order in chiral perturbation theory, including the effects of couplings to two nucleon currents \cite{Hoferichter:2017olk}. The result of this calculation can be incorporated in the cross section by taking $f_N = 0.308 \pm 0.018$ in Eq.~\eqref{eq:sig}, and we use this value in evaluating the experimental bounds.

In Figure~\ref{fig:thermal} we show the current direct detection constraints on the portal coupling arising from spin-independent scattering. At large DM masses these are dominantly from Xenon1T \cite{Aprile:2017iyp} and PandaX2 \cite{Tan:2016zwf}, while at $m_{\rm DM} \lesssim 5~\GeV$ CDMS \cite{Agnese:2017jvy} sets stronger limits (at even smaller DM masses, $m_{\rm DM} \lesssim 2~\GeV$, CRESST-III is relevant \cite{Petricca:2017zdp}). For clarity, we do not plot the uncertainties arising from the experimental set-up, astrophysics, and from $f_N$. These can alter the bound on $\lambda$ by a factor of $\sim 2$, which does not change the qualitative interpretation of the results. These bounds already rule out the minimal scalar Higgs portal model over the majority of parameter space, except for a region where resonant annihilation via the visible sector Higgs is possible, and a region at very large DM masses. In the latter case, values of the portal coupling $\gtrsim 1$ are required, close to the limit of perturbativity. We also plot the neutrino floor, discussed further in Section~\ref{sec:future}, which gives a lower limit on the DM scattering cross section that a conventional direct detection experiment can discover.
 
If $m_{\rm DM} < m_h / 2$, Higgs portal DM models can lead to a significant SM Higgs invisible decay rate, and collider searches for such processes can be relevant 
 \cite{Barger:2007im,Low:2011kp,Djouadi:2012zc}. The current constraint on the Higgs invisible width $\Gamma_{\rm invis}$ from the LHC is $\Gamma_{\rm invis}/ \left(\Gamma_{\rm invis} + \Gamma_{\rm SM}  \right) < 0.24$, where $\Gamma_{\rm SM}$ is the SM width, assuming that the Higgs production rate is given by the SM prediction \cite{Aad:2015pla,Khachatryan:2016whc}. This bound is dominantly from direct searches for invisible decays, which lead to a large missing momentum in combination with particular visible signatures, for example after associated vector boson Higgs production. Such limits are stronger than indirect constraints from the measurement of the decay rate of the Higgs to SM states. Using the prediction for the invisible Higgs decay rate in the Higgs portal model,
\begin{equation}
 \Gamma_{\rm invis} = \frac{\lambda^2}{32 \pi} \frac{v^2}{m_h} \sqrt{1- \frac{4 m_{\rm DM}^2}{m_h^2}} ~,
\end{equation}
where $v= 246.2~\GeV$, the corresponding bound on the portal coupling is plotted in Figure~\ref{fig:thermal}. Alone this would rule out scalar Higgs portal models with $m_{\rm DM} \lesssim 40~\GeV$, however given recent progress in direct detection searches it does not presently constrain any parameter space that is not otherwise excluded.\footnote{Direct detection bounds on models with low mass DM are weak, but these are ruled out by indirect detection constraints from Fermi and observations of the cosmic microwave background.} 

Future improvements in searches for invisible Higgs decays at the high luminosity LHC or an $e^+$ $e^-$ collider, discussed in Section~\ref{sec:con}, could be sensitive to couplings down to $\lambda \simeq 10^{-2.5}$. However, this will only constrain a small part of the viable parameter space in Figure~\ref{fig:thermal}, around the resonance region \cite{Han:2016gyy}. Even for fermion Higgs portal DM with pseudoscalar couplings, for which the direct detection bounds are extremely weak, almost all of the parameter space with $m_{\rm DM} < m_h/2$ is already excluded by the existing collider limits, and further improvements will only have a limited impact \cite{Escudero:2016gzx,Dutta:2017sod}. For DM  masses larger than $m_h/2$ other signatures at particle colliders are possible \cite{Chacko:2013lna,Craig:2014lda,Balazs:2017ple}.  In the minimal model that we consider these are far weaker than direct detection bounds, but they could be significant if the state that is coupled to the Higgs is a mediator to a dark sector, rather than the DM itself.

Higgs portal models can also lead to signals in indirect detection experiments, as a result of late time annihilations to the visible sector \cite{Okada:2013bna,Duerr:2015mva,Duerr:2015aka}. The Fermi experiment has searched for gamma ray emission from dwarf spheroidal galaxies, leading to constraints on the DM annihilation cross section evaluated at the typical present day DM velocity \cite{Ackermann:2015zua}. The null results can again be interpreted as bounds on $\lambda$ as a function of $m_{\rm DM}$ \cite{Duerr:2015mva}, and these are plotted in Figure~\ref{fig:thermal} (in parts of parameter space where they are comparable to direct detection constraints). The region just above the lowest point of the resonance is ruled out, since in this case the annihilation cross section in the present day universe is larger than in the early universe when the relic abundance is set, due to the smaller DM velocity today. Fermi has also observed a possible excess in gamma rays emitted from the galactic centre, which could be due to DM annihilations \cite{Berlin:2014tja,Daylan:2014rsa}. This has been studied in the context of the Higgs portal model that we consider \cite{Cuoco:2016jqt,Kim:2016csm,Sage:2016xkb}, but it can not be easily accommodated  \cite{Escudero:2016gzx}. To obtain a present day annihilation cross section that matches the tentative signal, while also being compatible with the measured DM relic density, a DM mass very close to the bottom of the resonance is needed. For slightly larger DM masses, a  present day annihilation cross section that matches the excess corresponds to a cross section in the early universe that is too small, leading to an over abundance of DM. For smaller DM masses, values of the portal coupling that lead to the observed relic abundance predict a too small annihilation cross section today.\footnote{The galactic centre might be easier to fit in  more complex Higgs portal models \cite{Wang:2014elb,Basak:2014sza,Casas:2017rww}.}

The currently viable region at large DM masses could be explored by the Cerenkov Telescope Array (CTA), which is currently in development \cite{Cline:2013gha,Silverwood:2014yza,Beniwal:2015sdl}. This is aimed at detecting gamma rays with energies between $\sim 10~\GeV$ and $\sim 100 ~\TeV$. However, the expected reach of CTA is strongly dependent on the DM density profile near the galactic centre, and this is currently subject to significant uncertainties. In Figure~\ref{fig:thermal} we show the projected sensitivity of CTA assuming the optimistic case of an NFW DM profile \cite{Silverwood:2014yza}, for which the DM density is relatively high at the galactic centre. If the DM is less dense in this region, for example as predicted by the Einasto profile \cite{Navarro:2003ew}, CTA will not be sensitive to any parameter space that is not already ruled out by direct detection experiments for the model we consider.

Late time annihilations can also result in a significant rate of position production in Higgs portal models, both for DM masses above and below the threshold for production of electroweak (EW) gauge bosons \cite{Urbano:2014hda}. As a result, measurements of the antiproton flux by AMS-02 \cite{Aguilar:2016kjl} constrain the value of the portal coupling \cite{Cui:2016ppb,Cuoco:2016eej}, and the bounds obtained can be comparable to those from direct detection experiments in the high DM mass region. However, these limits are subject to significant uncertainties, dominantly from the propagation of positrons in the galaxy. In Figure~\ref{fig:thermal} the constraints obtained using the  propagation model described in \cite{Cuoco:2016eej} are plotted. If other propagation models are assumed these results can differ by a factor of $\sim 3$ in the value of the portal coupling, and some propagation models lead to weaker bounds that are further inside the region already excluded by direct detection searches. There is a claimed excess in the position flux, corresponding to DM masses in the range $20~\GeV \lesssim m_{\rm DM} \lesssim 80~\GeV$. However, similarly to the Fermi galactic centre excess, this is hard to fit within the minimal Higgs portal model. Finally, models with small DM masses are constrained by indirect detection bounds from the cosmic microwave background (CMB) \cite{Slatyer:2015jla} and BBN \cite{Henning:2012rm}. For the minimal Higgs portal model these are weaker than the limits from searches for Higgs invisible decays.

In the remainder of this paper we show that changing the cosmological history of the universe can alter the value of the portal coupling required to obtain the measured DM relic abundance, and consequently can allow models with scalar Higgs portal DM to evade experimental constraints. In passing we note that this could also be achieved by introducing a more complex hidden sector. Adding extra hidden sector states can lead to new annihilation channels \cite{Casas:2017jjg}, which can reduce the relic abundance and permit smaller values of $\lambda$. Additionally, models containing two real hidden sector scalar fields with a softly broken U(1) symmetry coupled to the visible sector through a Higgs portal operator have been studied \cite{Barger:2008jx}, and can lead to viable parameter space. Sommerfeld enhancement of the DM annihilation cross section can also have an important effect in more complicated models with additional new light states, and can allow models with large DM masses \cite{MarchRussell:2008yu}, as well as potentially enhancing indirect detection signals \cite{Arina:2010an}.\footnote{It has recently been suggested that for sufficiently heavy DM, Sommerfeld enhancement from the SM Higgs itself could have a significant effect on the DM relic abundance \cite{Harz:2017dlj}.}

Fermion Higgs portal DM with a pseudoscalar coupling to the SM Higgs also remains viable over large parts of parameter space \cite{Escudero:2016gzx}. In this case, the direct detection cross section is velocity suppressed and extremely weak (and probably below the neutrino floor, depending on the present day DM velocity distribution). In contrast, its annihilation rate is not suppressed at low velocities, and future indirect detection searches could be relevant \cite{Escudero:2016gzx}. Meanwhile, in models with fermion DM with scalar couplings to the Higgs, the extra states required to UV complete the portal operator can have important effects \cite{Arina:2010wv}, and more complex Higgs sectors could lead to blind spots in direct detection searches \cite{Baum:2017enm}.

\section{Moduli and late time matter domination} \label{sec:nontherm}

Turning to the main focus of our work, we now consider theories that have a non-thermal cosmological history due to the presence of a relatively light scalar with a highly suppressed decay rate. If the scalar is displaced from the minimum of its potential, this typically leads to a period of late time matter domination, followed by entropy injection when it decays.  The possibility that late decaying particles could alter the universe's cosmological history has been studied extensively  \cite{Ellis:1984er,Ellis:1984eq,Scherrer:1984fd,Rajagopal:1990yx,Chung:1998ua,Moroi:1999zb}. Other possibilities for modifying the cosmological history include a period of kinetic energy domination, which also alters the DM relic abundance \cite{Salati:2002md}, but this is less motivated from a UV perspective.

\subsection{String moduli}

UV completions of the SM by string theory models typically include scalar moduli fields, which are associated to deformations of the size or shape of the extra spatial dimensions that are compactified to get a 4 dimensional low energy effective theory. The process of stabilising the extra dimensions, for example through non-perturbative effects, leads to the moduli getting masses. The spectrum of moduli depends on the details of a compactification, however the lightest is expected to have a mass that is parametrically of the same order as the gravitino mass. Additionally, in some string constructions the properties of the moduli can be calculated explicitly \cite{Acharya:2008bk,Acharya:2010af}.

Motivated by the success of gauge coupling unification when the SM is completed to the Minimal Supersymmetric Standard Model (MSSM) with soft masses not far above the EW scale (which also solves most of the hierarchy problem), we consider models in which the gravitino mass and the mass of the lightest modulus are relatively close to the $\TeV$ scale.\footnote{We assume that supersymmetry breaking is mediated to the visible sector by Planck suppressed operators. In models of gauge mediation, obtaining moduli masses that are large enough for a viable cosmological history in combination with visible sector soft masses at the $\TeV$ scale is challenging.} Such a model automatically leads to a period of late time matter domination.

The decay rate of a string modulus to fermions $\phi \rightarrow \bar{\psi}{\psi}$ is parametrically
\beq \label{eq:phidecay}
\Gamma_{\phi} = c \frac{m_{\phi}^3}{M_{\rm Pl}^2} ~,
\eeq
where $c$ is a constant that is expected to be of order 1 \cite{Acharya:2008bk}, and $M_{\rm Pl}$ is the reduced Planck mass.\footnote{Although this is a good approximation in many string models, the modulus decay rate could be dramatically different in compactifications with an exponentially large internal volume and modulus couplings enhanced by powers of the volume compared to Eq.~\eqref{eq:phidecay} \cite{Cicoli:2010ha}.} In many string models, the lightest modulus mass eigenstate is a mixture of multiple UV moduli, and in this case it often has relatively similar couplings to the DM and visible sector states. However, given that we do not commit to a particular underlying model, we assume that the lightest modulus might plausibly have a branching fraction to DM that is a factor $\sim 10^8$ smaller than to the visible sector.

\subsection{Non-thermal cosmological histories}

For simplicity we focus on models in which there is a single modulus $\phi$, with mass $m_{\phi}$, that is substantially lighter than the others. The final DM relic abundance is then determined by the dynamics of this modulus. More generally, string compactifications are typically expected to have multiple moduli with similar masses and decay rates, which could lead to interesting new dynamical regimes (we intend to return to this possibility in future work).

We assume that inflation occurs at a scale that is not extremely low, in particular that the Hubble scale during inflation $H_I \gtrsim m_{\phi}$. As a result the lightest modulus generically gets a vacuum expectation value (VEV) of order $\left<\phi \right> \sim M_{\rm Pl}$ \cite{Acharya:2008bk}, since around this value higher dimension operators in its potential become significant. The modulus' subsequent evolution is given by
\beq \label{eq:phiev}
\ddot{\phi} + 3 H \dot{\phi} + m_{\phi}^2 \phi =0 ~,
\eeq
where $H$ is the Hubble parameter. Consequently, the modulus remains fixed at its initial VEV until the Hubble parameter drops to $H_{\rm osc} \simeq m_{\phi}$, when it begins oscillating around the (zero temperature) minimum of its potential. 

The coherent oscillations of the modulus act a non-relativistic matter contribution to the energy density of the universe, and after the modulus begins oscillating its energy density is diluted as $\sim 1/a\left(t\right)^3$, where $a\left(t\right)$ is the scale factor of the universe. In contrast, the energy density in the radiation bath decreases as $\sim 1/a\left(t\right)^4$. Since the modulus begins oscillating when its energy density, $\rho_{\phi}\simeq m_{\phi}^2 \left<\phi\right>^2 \sim m_{\phi}^2 M_{\rm Pl}^2$, is comparable to the energy in the radiation bath, $\rho_{\rm rad} \sim \sim H_{\rm osc}^2 M_{\rm Pl}^2$, it quickly comes to dominate the universe. During the subsequent matter dominated era the Hubble parameter is approximately $H^2 = \rho_{\phi}/\left(3M_{\rm Pl}^2 \right)$, where $\rho_{\phi}$ is the energy density in the modulus, and the universe's scale factor increases with time as $a(t) \sim t^{2/3}$. 

Matter domination continues until the Hubble parameter drops to be comparable to the modulus decay width, when the majority of the moduli quanta decay. We follow the convention of defining the reheating temperature $T_{\rm RH}$ such that $\Gamma_{\phi} = H\left(T_{\rm RH}\right)$, which is approximately the temperature  when  the universe passes back into radiation domination. From Eq.~\eqref{eq:phidecay},
\beq
T_{\rm RH} = \left(\frac{90}{\pi^2 g\left(T_{\rm RH}\right)} \right)^{1/4} m_{\phi} \sqrt{\frac{c m_{\phi}}{M_{\rm Pl}}}~,
\eeq
where $g\left(T_{\rm RH}\right)$ is the effective number of relativistic degrees of freedom at reheating.

A lower bound on the mass of the lightest modulus comes from the observational requirement that it decays before BBN, so that the universe is radiation dominated during this process. This corresponds to a reheating temperature  $T_{\rm RH} \gtrsim 10~\MeV$, and a modulus mass $\gtrsim 10~\TeV$, for $c$ of order $1$ in Eq.~\eqref{eq:phidecay} (the precise lower bound on $T_{\rm RH}$ does not significantly modify the possible DM phenomenology).

Although matter domination continues until relatively late times, the modulus begins decaying as soon as it starts oscillating. These early decays only transfer a small fraction of energy stored in the modulus field to radiation (or DM) until $H$ drops to $\sim \Gamma_{\phi}$. However, they still have an important impact on the system's dynamics and the final DM yield. 
If there is no energy in the radiation bath when the modulus first starts oscillating, the early decays mean that the maximum temperature reached by the bath is of order $T_{\rm max} \sim m_{\phi}$. This occurs when $t/t_{\rm osc} \sim \mathcal{O}\left(1\right)$, where $t_{\rm osc}$ is the time when the modulus begins oscillating \cite{Giudice:2000ex}, soon after the onset of matter domination and long before radiation domination re-emerges. Subsequently, the temperature of the radiation bath drops as $T \sim a^{-3/8}$, and therefore $H \sim T^{4}$, until the majority of the modulus quanta have decayed. 

Instead, in a simple model with a single modulus and a period of radiation domination before the modulus begins oscillating, there are two contributions to the energy density of the radiation bath. First, there is energy previously in the bath before matter domination, and this is supplemented by new energy produced from modulus decays. 
As a result, the temperature of the universe never increases during the modulus' decay \cite{Scherrer:1984fd}. Although the energy from primordial radiation can be greater than that from modulus decays until relatively late times, it is always eventually subdominant.\footnote{Once it begins oscillating, the energy density in the modulus is greater than in primordial radiation due to redshifting, and almost all of the energy density in the modulus is eventually converted back to radiation once the modulus completely decays.} Further, we find from a numerical analysis that in all of the parameter space of interest the final DM relic density is independent of whether there is any energy in the radiation bath when the modulus starts oscillating or not. That is, the dynamics that set the DM relic abundance happen after the primordial energy in radiation is sufficiently redshifted that it is irrelevant compared to energy from the modulus decay, and these dynamics are independent of the prior conditions (this matches the conclusion in \cite{Drees:2017iod}). The multiple moduli that are expected in typical string models will result in an even longer period of matter domination, which will make any initial energy in the radiation bath even less significant.

As well as string moduli, we could also consider theories containing weakly coupled scalars that lead to a period of matter domination, but which decay with a different parametric rate to Eq.~\eqref{eq:phidecay}. For example, non-thermal cosmologies have also been proposed to arise due to the dynamics of the saxion (the scalar counterpart of the axion) in F-theory models, and this can have a decay rate that is suppressed by the axion decay constant \cite{Heckman:2008jy}. However, for a given final reheating temperature, the viable DM parameter space will be similar to that in a model with a string modulus. The primary change is that a given reheating temperature corresponds to a different saxion/ modulus mass, due to the modified decay rate. As a result, a particular energy density corresponds to a different saxion/ modulus number density, so that, in some parts of parameter space, the branching fraction into DM (discuss in more detail shortly) that leads to a particular DM relic abundance will change.

\section{Higgs portal DM with a non-thermal cosmological history} \label{sec:hpnon}

A non-thermal cosmological history can dramatically alter the dynamics of the DM in the early universe and its relic abundance, compared to theories with a thermal history. During matter domination DM can be produced directly from the decays of the moduli, parameterised by the modulus branching fraction to DM, defined as \footnote{DM production from decays has been studied in, for example, \cite{Jeannerot:1999yn,Fujii:2002kr,Fujii:2001xp,Lin:2000qq,Bi:2009am,Dutta:2009uf}.}
\begin{equation}
B = \frac{\Gamma_{\phi \rightarrow {\rm DM}}}{\Gamma_{\phi \rightarrow {\rm DM}} + \Gamma_{\phi \rightarrow {\rm vis}} } ~,
\end{equation}
where $\Gamma_{\phi \rightarrow {\rm DM}}$ and $\Gamma_{\phi \rightarrow {\rm vis}}$ are the modulus decay rates to DM and the visible sector respectively. If the temperature is high enough $T \gtrsim m_{\rm DM}$, interactions between visible sector states in the thermal bath can also produce DM. Meanwhile, if its number density becomes sufficiently large, DM annihilations to the visible sector will be relevant. Additionally,  entropy is injected into the system as long as modulus decays continue, which acts to decrease the DM yield. 

The possible dynamics of Higgs portal DM models in theories with a non-thermal cosmological history are similar to those of WIMP DM candidates, which have been classified in \cite{Kane:2015qea} (WIMP DM models with a more complex hidden sector have also been studied \cite{Acharya:2017kfi}). One difference is that the annihilation cross section of Higgs portal DM has an unusual temperature dependence, due to new channels opening up as mass thresholds are passed and the possibility of resonant annihilation.

We calculate the final DM yield by numerically integrating the energy density in the modulus, radiation, and DM during the evolution of the universe. For definiteness, we fix the coefficient $c=1$ in Eq.~\eqref{eq:phidecay} (this does not significantly change the phenomenology) giving a definite relation between the modulus mass and the reheating temperature. A model is therefore specified by four parameters: the portal coupling, the DM mass the, modulus branching fraction to DM, and the reheating temperature. The equations determining the evolution of the DM yield are well known \cite{Giudice:2000ex}, and it is convenient to define a dimensionless scale factor $A = a T_{\rm RH}$. Further, we work in terms of dimensionless variables
\beq
\begin{aligned} \label{eq:cmdef}
\Phi & = \frac{A^3}{T_{\rm RH}^4} \rho_{\phi} ,\qquad X & = \frac{A^3}{T_{\rm RH}^3} n_{\chi} , \qquad 
R & = \frac{A^4}{T_{\rm RH}^4} \rho_{R} ,
\end{aligned}
\eeq
where $n_{\chi}$ is the DM number density, and $\rho_{\phi}$ and $\rho_{R}$ are the energy densities in moduli and radiation respectively. With this definition $X$ remains constant when no DM states are produced or destroyed, even when entropy in injected into the universe by modulus decays.

In Appendix~\ref{sec:ap} we give expressions for $\frac{d\Phi}{dA}$, $\frac{d X}{dA}$, $\frac{d R}{dA}$, and $\frac{d T}{dA}$, which reproduce those in  \cite{Drees:2017iod}. $\Phi$ decreases with time due to modulus decays, $R$ increases due to modulus decays and can also be slightly altered by energy transfer to or from the DM. $X$ is increased by modulus decays and increased or decreased by production from, or annihilation to, the thermal bath depending on the sign of $X_{\rm eq}^2 - X^2$, where $X_{\rm eq}$ is its equilibrium value at a given temperature. The number of relativistic degrees of freedom changes as the temperature of the universe drops during the modulus decay, and this can affect the DM relic abundance by a factor of $\mathcal{O}\left(1\right)$. We include the effect of this following  \cite{Drees:2017iod} (around $\Lambda_{\rm QCD}$ the number of degrees of freedom can be obtained from the equation of state \cite{Drees:2015exa}).

We primarily consider models with DM masses $\gtrsim 100~\MeV $, since obtaining the measured DM relic abundance for smaller masses requires that the portal coupling either has a tiny value, which would be challenging to observe experimentally, or a large value, which is already experimentally excluded. For simplicity, we use the DM annihilation cross section to SM quarks, leptons, gauge bosons, and the Higgs, described in Section~\ref{sec:hp}. For DM masses below the QCD scale, the cross section for annihilations to hadrons will be a source of major uncertainty, although for $100 \lesssim m_{\rm DM} \lesssim \GeV$ a significant fraction of annihilations are to muons and this is relatively unaffected by QCD corrections. An accurate calculation of the relic abundance in this DM mass range would require both the non-perturbative annihilation cross section to hadrons, and the inclusion of finite temperature effects due to the  QCD plasma, which could plausibly change the required value of the portal coupling by an order of magnitude. However, given that we do not expect our simple model to precisely reproduce UV complete theories, since these are expected to contain multiple moduli, this approximation for the annihilation cross section is sufficient for our current purposes.

\subsection{Regimes of DM in non-thermal cosmologies} \label{sec:reg}

Depending on the reheating temperature and the modulus branching fraction to DM, the DM can have different dynamics. The resulting relic abundance can be larger or small than would occur in a model with the same DM properties and a thermal cosmological history. 

Given the results in Figure~\ref{fig:thermal}, for Higgs portal models to be viable over large parts of parameter space the DM relic abundance must be reduced relative to that in a thermal cosmology. This can happen in different ways and, although for our final results we will rely on a numerical integration of the Boltzmann equations, the approximate dependence of the relic abundance on the model parameters can be understood analytically.

\begin{itemize}
 \item {\bf DM production from the thermal bath, without chemical equilibrium} 
 
In this regime DM is dominantly produced from scatterings of states in the visible sector thermal bath. If the portal coupling is sufficiently small, the DM number density will never be large enough for annihilations to become significant, and chemical equilibrium is not reached. Production becomes inefficient once the visible sector thermal bath temperature drops below $\sim m_{\rm DM}$. Subsequently the comoving DM number density $X$ remains constant, and the DM yield $n_{\chi}/s$ (where $s$ is the  entropy density) is reduced by entropy injection from the modulus decays.

In this case, most of the DM is produced when the visible sector temperature is $T_{\rm pro}\sim m_{\rm DM}$. At earlier times when the temperature $T > T_{\rm pro}$, a Hubble time is shorter by a factor $\sim T_{\rm pro}^4 / T^{4} $ and any DM produced at this point is diluted by an extra factor of $ a\left( T_{\rm pro}\right)^3 /a\left(T\right)^3 \sim T_{\rm pro}^8 / T^{8}$ due to the expansion of the universe, although the DM production rate is enhanced by the increased number density of visible sector states $n_{\rm vis}\left(T\right) /n_{\rm vis}\left(T_{\rm pro}\right) \sim T^3 / T_{\rm pro}^{3} $. Combining these factors, DM is dominantly created from the thermal bath at the lowest kinematically allowed temperatures. This remains the case even once they strong temperature dependence of the cross section in Higgs portal models is included.\footnote{Freeze in via a renormalisible operator is also IR dominated in a radiation dominated universe, and the stronger dependence of the Hubble parameter on the temperature during matter domination favours low temperatures even more strongly.}

The  dependence of the DM relic abundance on a model's parameters can be estimated from the DM production rate at $T_{\rm pro}$. This is $dn_{\chi}/dt \sim n^2 \left<\sigma v\right>$, where $n \sim T_{\rm prod}^3$ is the number density of SM states,   $\left<\sigma v\right>$ is the DM annihilation cross section evaluated at $T_{\rm pro}$, and we have ignored numerical factors that are $\lesssim 100$. As a result the DM yield at the time of production is \beq
y\left(T_{\rm pro} \right) = \frac{n_{\chi}}{s} \sim \frac{1}{\frac{2\pi^2}{45} g m_{\rm DM}^3} \frac{dn_{\chi}}{dt} \frac{1}{H} ~,
\eeq
and during the matter dominated era $H \sim \frac{T^{4}}{T_{\rm RH}^2 M_{\rm Pl}}$. Subsequently, the DM yield is decreased by a factor of $y\left(T_{\rm RH}\right)/y\left(T_{\rm pro} \right)  = \left(T_{\rm RH}/m_{\rm DM} \right)^5$ due to entropy injection, since $n_{\chi}\left(T_{\rm RH}\right)/n_{\chi}\left(T_{\rm pro}\right) = a(T_{\rm pro})^3/a(T_{\rm RH})^3$. Therefore, the final DM yield is approximately
\beq
y_{\infty} \sim 10^{-9} m_{\phi}^{21/2} M_{\rm Pl}^{-5/2} m_{\rm DM}^{-6} \left<\sigma v \right> ~.
\eeq
In calculating the viable DM parameter space, the strong temperature dependence of $\left<\sigma v\right>|_{T_{\rm pro}}$ on $m_{\rm DM}$ as mass thresholds are crossed is important. However, we note that if $\left<\sigma v\right>$ was temperature independent, the DM relic abundance would be proportional to $\sim T_{\rm RH}^7 M_{\rm Pl} m_{\rm DM}^{-5} \left<\sigma v \right> $ in this regime. As a result, contours of constant DM relic abundance with $T_{\rm RH}$ fixed would correspond to portal couplings that varied with the DM mass as $\lambda \sim m_{\rm DM}^{5/2}$ (dropping a dimensionful constant).

\item {\bf Production from thermal bath, with chemical equilibrium and freeze out during matter domination}

If the portal coupling is sufficiently large, production from the SM thermal bath is efficient enough that annihilations become significant and the DM reaches chemical equilibrium. This persists until the temperature of the thermal bath drops to  $\sim m_{\rm DM}$, when the equilibrium DM number density drops rapidly. As in normal freeze out in a radiation dominated universe, the DM number density initially decreases, until annihilations become slow compared to the Hubble parameter and the DM falls out of chemical equilibrium. Subsequently the DM yield is decreased by entropy injection from the modulus decays, which reduces its final relic abundance compared to if freeze out had occurred with a thermal cosmological history.

Freeze out takes place when the Hubble parameter is roughly $H \sim m_{\rm DM}^4 / m_{\phi}^3$. As a result, DM annihilations remain efficient until the DM number density drops to $n_{\chi} \sim  m_{\rm DM}^4 / \left( m_{\phi}^3 \left<\sigma v \right> \right)$. Including the extra entropy injection after freeze out, the final DM yield is approximately 
\beq
\begin{aligned}
y_{\infty} & \sim 0.01 m_{\phi}^{9/2} m_{\rm DM}^{-4} M_{\rm Pl}^{-5/2} \left<\sigma v \right>^{-1} \\
& \sim 0.01 T_{\rm RH}^3 m_{\rm DM}^{-4} M_{\rm Pl}^{-1} \left<\sigma v \right>^{-1} ~.
\end{aligned}
\eeq
As before, the strong temperature dependence of $\left<\sigma v \right>$ is crucial. If $\left<\sigma v \right>$ was temperature independent, contours of fixed DM relic abundance with $m_{\phi}$ constant would correspond to portal couplings that vary as $\lambda \sim m_{\rm DM}^{-3/2}$.

\item {\bf Production from modulus decay without reaching chemical equilibrium}

If the modulus branching fraction to DM is sufficiently large, direct decays can be the main source of DM. In this case, the majority of the DM is produced when $T = T_{\rm RH}$, when most of the moduli decay. If the DM number density produced is not large enough for annihilations to be significant then $n_{\chi} \simeq B \rho_{\rm RH}/m_{\phi}$ at this time, where $\rho_{\rm RH} \simeq \frac{\pi^2}{30} g T_{\rm RH}^4$ is the modulus energy density when $T \simeq T_{\rm RH}$ (at which point $\rho_{\phi} \sim \rho_{\rm rad}$). 
The final DM yield when radiation domination begins is therefore
\beq
y_{\infty} \sim 0.2 B \sqrt{\frac{m_{\phi}}{M_{\rm Pl}}} ~.
\eeq
Contours of constant DM relic abundance are independent of $\lambda$ in this regime, and obtaining the measured DM relic abundance requires a modulus branching fraction 
\beq
B \simeq 10^{-3} \left(\frac{\GeV}{m_{\rm DM}} \right) \left(\frac{\GeV}{T_{\rm RH}} \right)^{1/3} ~.
\eeq

\end{itemize}

There are also other possible DM dynamics in non-thermal models. These do not reduce the DM relic abundance compared to models with a thermal cosmological history, and  are therefore less phenomenologically interesting for our present work.
\begin{itemize}
\item {\bf Production from modulus decay, with chemical equilibrium}

If the modulus branching fraction to DM is sufficiently large, the DM is dominantly produced from decays and it can reach chemical equilibrium. This leads to a quasi-static equilibrium, with production balanced by annihilations. Once the modulus has completely decayed DM production stops, and the DM continues annihilating until its number density drops to $n_{\rm DM} \simeq H/ \left<\sigma v \right>  \sim T_{\rm RH}^2 / \left( M_{\rm Pl} \left<\sigma v \right>  \right) $ \cite{Acharya:2009zt}. Since we are interested in parts of parameter space for which $T_{\rm RH} \ll m_{\rm DM}$, this results in a larger relic abundance than the thermal freeze out prediction.\footnote{If $T_{\rm RH} \sim m_{\rm DM}$ the relic abundance is set by freeze out in a radiation dominated universe.}

\item {\bf Freeze out after reheating} 

If its mass is sufficiently small, the DM can be in chemical equilibrium with the visible sector immediately after reheating, when the universe returns to radiation domination. In this case the DM remains in chemical equilibrium as long as $n_{\chi} \left<\sigma v \right> \gtrsim H$, and when this is violated freeze out occurs. The DM relic density is the same as if the universe had been radiation dominated throughout its history, reproducing the parameter space studied in Section~\ref{sec:hp}.
\end{itemize}

\begin{figure}
\begin{center}
 \includegraphics[width=0.485\textwidth]{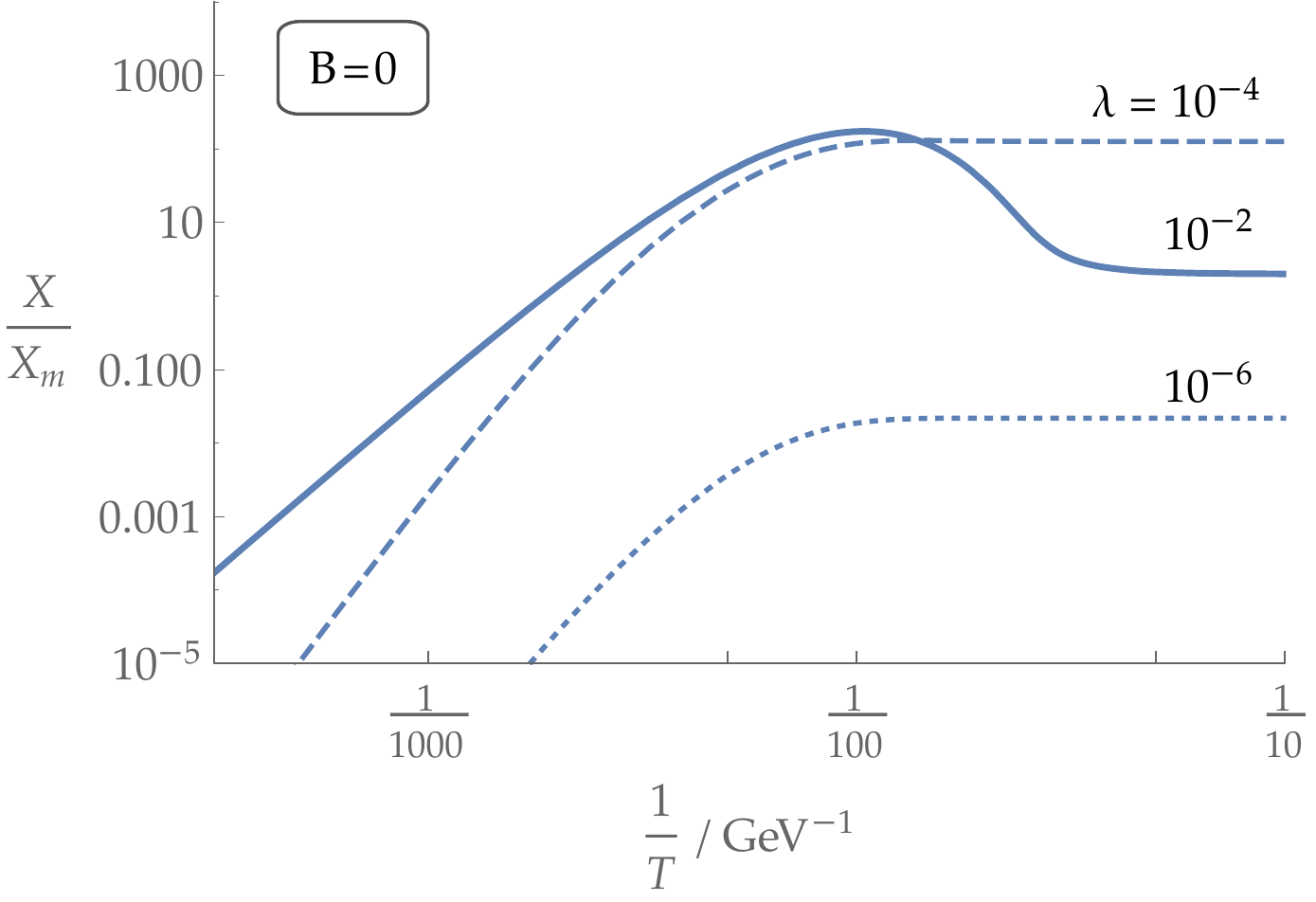}
 ~
 \includegraphics[width=0.485\textwidth]{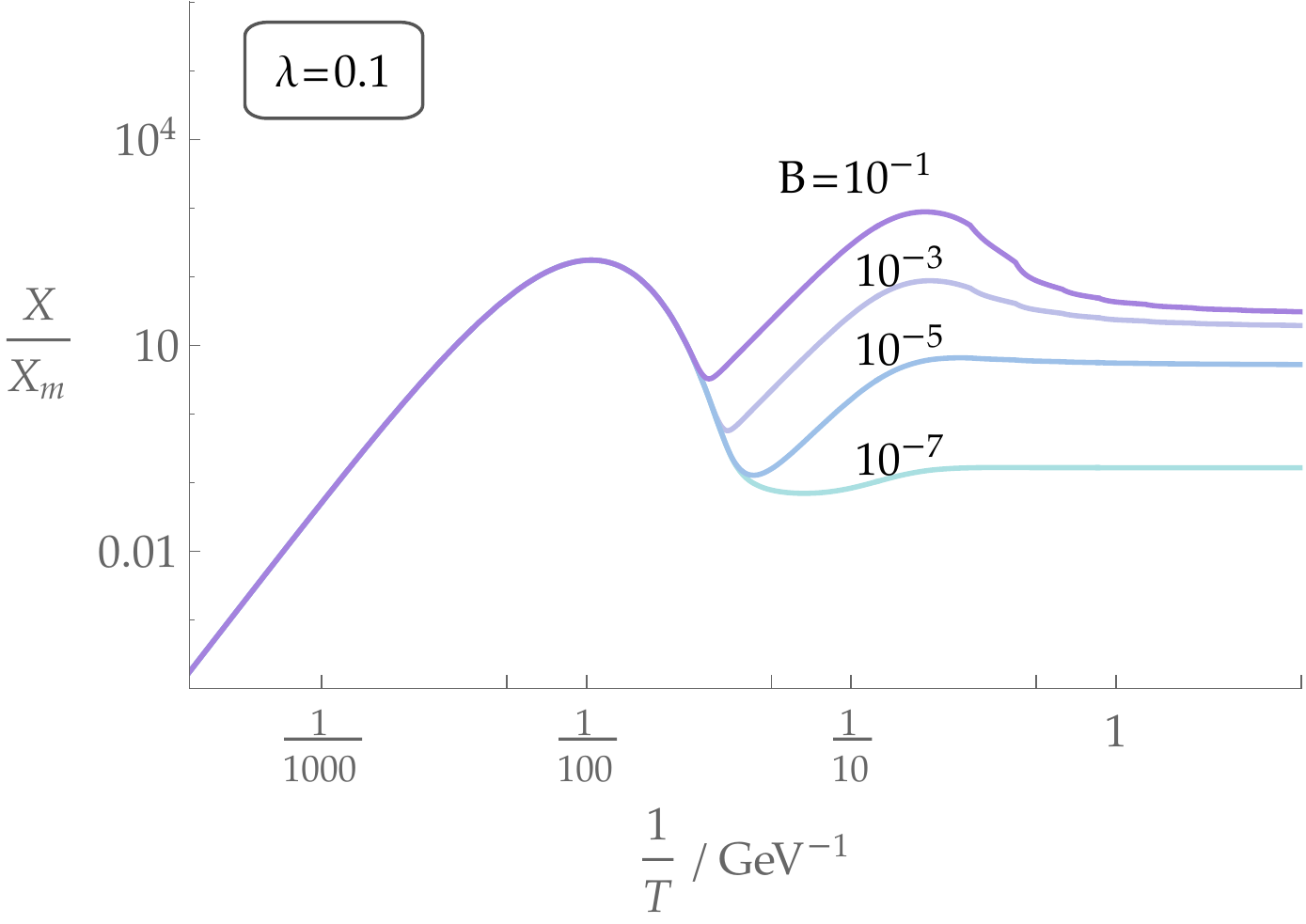}
\caption{{\bf \emph{Left:}} The evolution of the comoving DM number density $X$ (defined in Eq.~\eqref{eq:cmdef}, normalised to the value required to match the observed DM relic abundance $X_{\rm m}$) as a function of the visible sector temperature $T$, in non-thermal Higgs portal DM models with different values of the portal coupling $\lambda$. The reheating temperature is fixed to $10~\GeV$, the DM mass is set at $600~\GeV$, and the modulus branching fraction to DM is $B=0$, so that DM is only produced from the visible sector thermal bath.  {\bf \emph{Right:}} The evolution of $X$ in models with different values of the modulus branching ratio to DM. The reheating temperature is fixed to $10~\GeV$, the DM mass is $600~\GeV$, and the portal coupling is  $\lambda=0.1$. At early times DM is produced from the thermal bath, and then freezes out during matter domination. Subsequently, the majority of DM is produced from modulus decays, when the visible sector temperature is around the reheating temperature. }
\label{fig:densitywitht} 
\end{center} 
\end{figure}

We plot the comoving DM number density $X$ as the temperature of the universe drops (that is, as time progresses) for different non-thermal Higgs portal DM models in Figure~\ref{fig:densitywitht}. These results are obtained by numerically integrating the Boltzmann equations, for a fixed reheating temperature $T_{\rm RH} = 10~\GeV$ (or equivalently having taken $c=1$ in Eq.~\eqref{eq:phidecay}, a modulus mass of $\simeq 10^4~\TeV$) and a fixed DM mass of $600~\GeV$.\footnote{For clarity, we set the energy in the radiation bath before matter domination to zero in these plots. As discussed, this does not affect the final DM abundance.} 

In Figure~\ref{fig:densitywitht} left, we show models with zero modulus branching fraction to DM, so that production is purely from the thermal bath. When the portal coupling is small the DM never reaches chemical equilibrium, and increasing the coupling simply leads to more DM being produced, with the relic abundance proportional to $\lambda^2$. At a critical value of the coupling (in this case $\lambda \sim 10^{-3.5}$) the DM reaches chemical equilibrium. Increasing the portal coupling further decreases the relic abundance, since this is now set by freeze out during matter domination. As a result, there are two values of the portal coupling that lead to the observed DM relic abundance for this particular DM mass and reheating temperature, if $B=0$. 

In Figure~\ref{fig:densitywitht} right, we plot models in which the DM relic abundance is dominantly produced by modulus decays. At early times DM is produced from the thermal bath and reaches chemical equilibrium, before freezing out when the temperature  drops below $m_{\rm DM}$. Subsequently, modulus decays continue and most of the DM is produced when the temperature is $\sim T_{\rm RH}$. For $B\lesssim 10^{-3}$ the DM number density does not become large enough for annihilations to be efficient at late times, and the relic abundance is proportional to $B$. For larger values of the modulus branching fraction, DM annihilations become efficient, and a quasi-static equilibrium is reached. In this regime the final DM abundance is approximately independent of $B$.

\subsection{The parameter space for Higgs portal DM}\label{sec:ps1}

The parameter space for which Higgs portal DM makes up the full observed relic abundance can be found by scanning over different models, numerically integrating the Boltzmann equations for each. In particular, we fix the reheating temperature and modulus branching fraction since these are determined by the underlying string theory, and study how the relic density constraint translates into contours in the plane of portal coupling vs. DM mass. 

\begin{figure}
\begin{center}
 \includegraphics[width=0.485\textwidth]{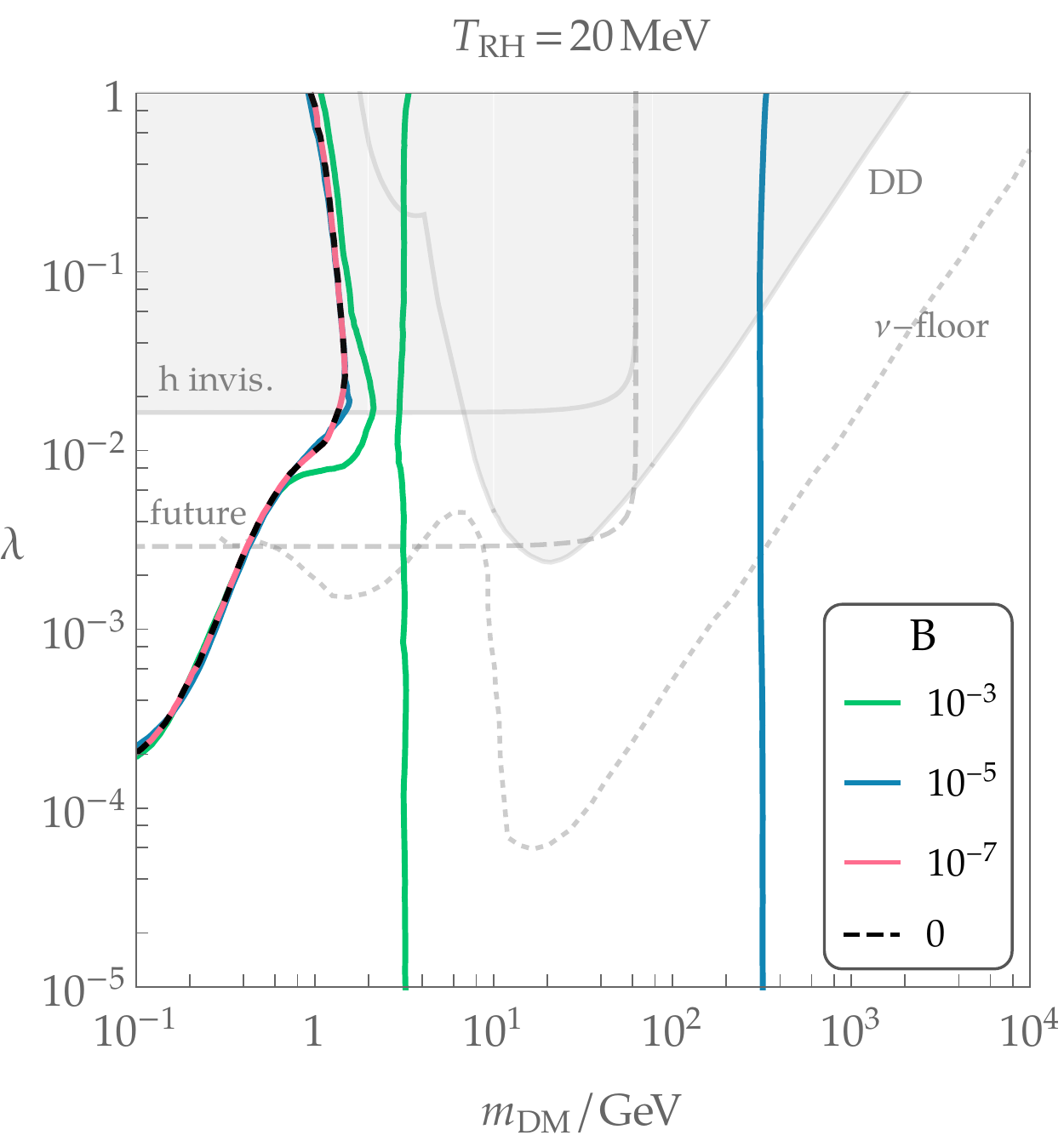}
 ~
 \includegraphics[width=0.485\textwidth]{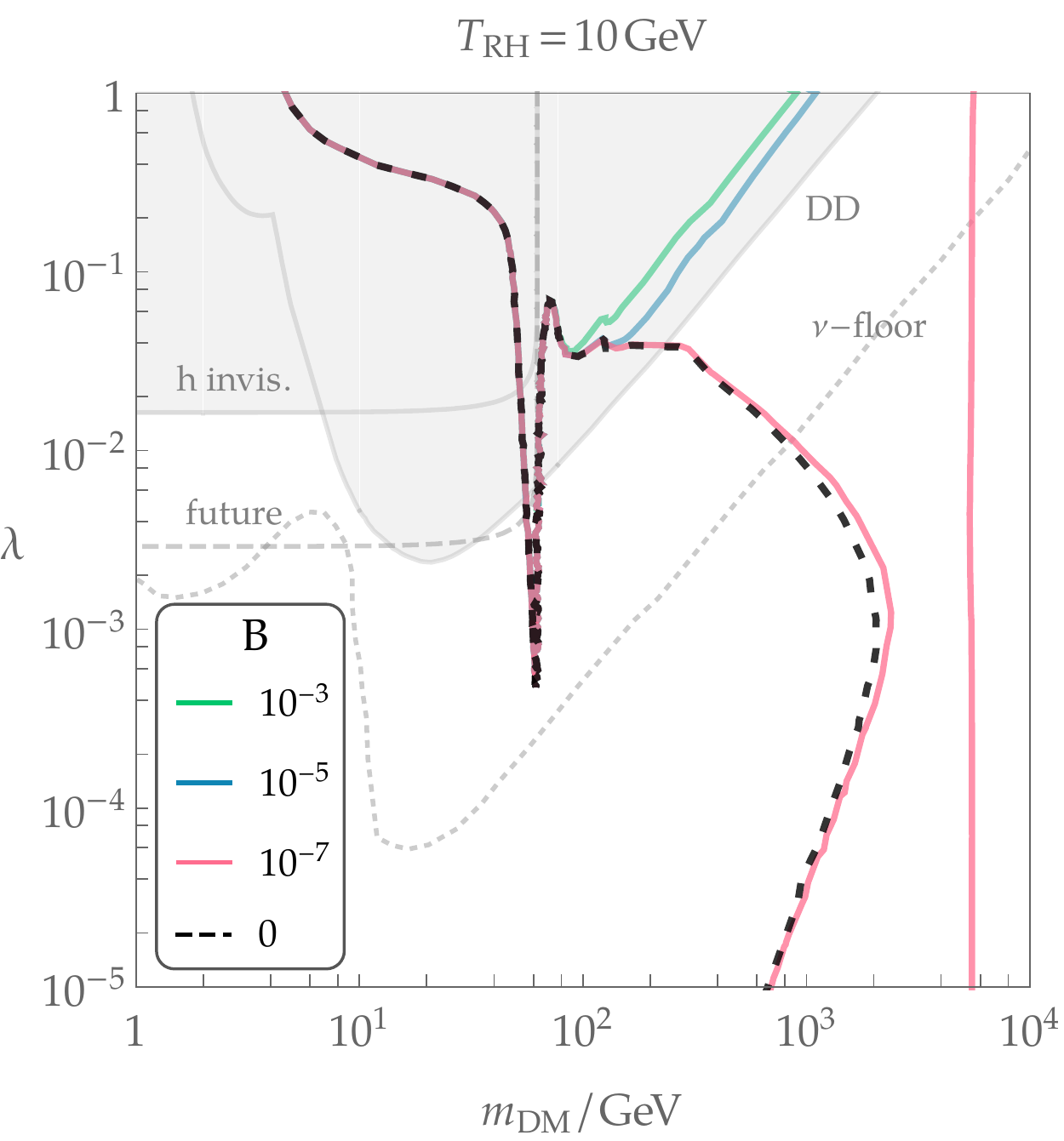}
\caption{{\bf \emph{Left:}} The parameter space for Higgs portal DM in models with a period of late time matter domination and a reheating temperature of $20~\MeV$. The contours show the values of the DM mass and portal coupling $\lambda$ that lead to the full measured DM relic abundance, for different values of the modulus branching fraction to DM $B$. For small DM masses production is dominantly from the visible sector thermal bath, and chemical equilibrium is reached if $\lambda \gtrsim 10^{-2}$. At larger masses, DM is mainly produced directly from modulus decays. The current constraints from LHC searches for invisible Higgs decays (h invis.) and direct detection bounds (DD) are shown shaded gray. The future reach of searches for invisible Higgs decays at the high luminosity LHC or an $e^+$ $e_-$ collider, sensitive to an invisible branching fraction of $1\%$, (future) is shown dashed gray, and the neutrino floor ($\nu-{\rm floor}$) is plotted with a dotted gray line. {\bf \emph{Right:}} The same parameter space for a model with a reheating temperature of $10~\GeV$. In this case production from the thermal bath can lead to viable models with relatively heavy DM if the modulus branching fraction to DM is small (chemical equilibrium is reached in such models if $\lambda \gtrsim 10^{-3}$). For even heavier DM, production directly from modulus decays can lead to the measured DM relic abundance for small values of $B$, while for $m_{\rm DM} \lesssim 200~\GeV$ the DM is in thermal equilibrium after reheating, and the required value of $\lambda$ is the same as in models with a thermal cosmological history.}
\label{fig:nonthermal} 
\end{center} 
\end{figure}

In Figure~\ref{fig:nonthermal} we plot the results for theories with a low reheating temperature $T_{\rm RH} = 20~\MeV$, close to the minimum allowed by  BBN, and for models with a higher reheating temperature $T_{\rm RH} = 10~\GeV$ (which is still low enough that the non-thermal history has an impact). The former corresponds to a modulus mass $\sim 100~\TeV$ and the latter  $\sim 10^4~\TeV$.\footnote{If the visible sector soft masses are similar to the modulus mass, this range can lead to a Higgs mass of $125~\GeV$ in the MSSM.} The required values of the portal couplings as a function of the DM mass are plotted for modulus branching fractions between $B = 10^{-3}$ and $B=0$, since allowing $B>10^{-3}$ does not open up any additional parameter space. The observed DM relic abundance can arise from models that are in all of the dynamical regimes discussed in Section~\ref{sec:reg}. Additionally, the gradients of the contours of the portal coupling vs DM mass match the approximate analytic calculations in Section~\ref{sec:reg}  reasonably closely, and even the numerical factors are fairly accurate.

Contours that are almost vertical correspond to models in which DM is produced directly from modulus decays without reaching chemical equilibrium. For very small modulus branching fractions to DM, production directly from  decays can lead to viable models with $m_{\rm DM} \gg ~\TeV$. Such heavy DM is relatively rare in theories with a thermal cosmological history, due to the large annihilation cross sections that are required assuming freeze out during radiation domination. Because the value of the portal coupling is unimportant in this regime, existing constraints from direct detection and collider searches can be evaded simply by taking $\lambda$ to be small (although given that the portal coupling plays no role in setting the DM relic density, the motivation for such models to be within reach of future experiments is also diminished).

For models with $T_{\rm RH} = 20~\MeV$, DM with a mass in the range $0.1~\GeV \lesssim m_{\rm DM} \lesssim \GeV$, and a relatively small value of the portal coupling, is produced from the visible sector thermal bath and does not reach chemical equilibrium, provided that $B \lesssim 10^{-2}$. In this part of parameter space contours of constant relic abundance have positive gradient. In models with $T_{\rm RH} = 10~\GeV$ the modulus branching fraction to DM must be relatively small in order to be in this regime, and it is possible for $600~\GeV \lesssim m_{\rm DM} \lesssim 2\TeV$ and $B \lesssim 10^{-6}$  (for larger values of $B$ production from direct modulus decays becomes more important than production from the thermal bath). Meanwhile, for larger values of the portal coupling in these mass ranges DM annihilations are important, and the relic abundance is set by freeze out during matter domination. In this part of parameter space the contours have negative gradient. In both of these regimes, the measured relic abundance occurs for smaller values of $\lambda$ than in theories with a thermal cosmological history.

Sufficiently light DM will be in chemical equilibrium with the visible sector thermal bath at the time of reheating, and will therefore freeze out during radiation domination, provided that the portal coupling is not tiny. For example, this is the case for the contours leading to the required relic abundance with $T_{\rm RH} = 10~\GeV$ and DM masses $\lesssim 200~\GeV$. In this regime the DM relic abundance is very close to that obtained from a model with a thermal cosmological history, plotted in Figure~\ref{fig:thermal}. 

Finally, for $T_{\rm RH} = 10~\GeV$ and $m_{\rm DM} \gtrsim 200~\GeV$, a quasi-static equilibrium is reached for $B \gtrsim 10^{-5}$ and $\lambda \gtrsim 0.1$. This increases the DM relic abundance compared to models with a thermal cosmological history, so that obtaining the measured DM abundance requires larger values of $\lambda$, which are already ruled out.

By varying the reheating temperature upwards from the minimum value allowed by BBN, the required DM relic abundance can arise for any value of the portal coupling if $m_{\rm DM} \gtrsim \GeV$ and $B=0$. Models with a particular non-zero modulus branching fraction to DM are also possible over the majority of this region, either through direct production from modulus decays, or by production from the thermal bath. Additionally, there is a significant part of parameter space with $100~\MeV \lesssim m_{\rm DM} \lesssim \GeV$, even for models with a relatively large modulus to DM branching ratio.

It can now be seen that calculating the relic abundance assuming that the DM is at the same temperature as the visible sector  does not lead to significant inaccuracies. Over most of the parameter space in which the DM is produced from the visible sector thermal bath the portal coupling is large enough for thermalisation, and even if not the DM states produced automatically have approximately the same temperature as the SM. In parts of parameter space in which the DM is produced by direct modulus decays and reaches a quasi-static equilibrium, the portal coupling is large enough that it will thermalise with the visible sector. Meanwhile, if the DM is dominantly produced from modulus decays and does not efficiently annihilate, it could remain at a higher temperature than the visible sector if the portal coupling is extremely small, so that scattering off the SM thermal bath is inefficient. This might have an impact on structure formation, discussed in Section~\ref{sec:future}, however it will not alter the DM relic abundance.

We also note that the contribution from the Higgs VEV to the DM mass, $m_{\rm DM}^2 \supset \lambda^2 v^2$, is not sufficiently large that the DM mass must be fine tuned to be small in any of the viable parameter space.\footnote{Although since the DM is a scalar its mass is quadratically sensitive to any new higher energy scales that it is sufficiently strongly coupled to.}  In models with $T_{\rm RH} = 20~\MeV$, in some parts of parameter space this contribution is comparable to the total DM mass, and it could even be the sole source of the DM mass (elsewhere it is significantly smaller than the total DM mass). Meanwhile, for models with $T_{\rm RH} = 10~\GeV$ it is negligible.

\section{Prospects for future observation and phenomenology} \label{sec:future}

Since the viable parameter space of non-thermal Higgs portal DM models extends over a large range of DM masses and values of the portal coupling, far beyond the current constraints, it is interesting to consider the possibilities for future detection. 

Observation of a signal that is challenging to obtain from simple models assuming a thermal cosmological history could point towards either a non-thermal cosmology or a more complex hidden sector \cite{Bertone:2017adx}. This would be the case if, for example, DM with a relatively small mass $\lesssim \GeV$ is detected. Another interesting possibility is that the DM could be observed in multiple different types of experiment. For example, if the DM scattering cross section measured in a direct detection experiment matched that predicted by Higgs portal models after a collider observation of invisible Higgs decays, this would be strong evidence in favour of such a theory. We will see that this combination of observations could plausibly occur given the potential reach of future experiments.\footnote{Higgs portal DM could also affect the SM EW phase transition \cite{Ghorbani:2018yfr}.}

\subsection{Collider searches}

As discussed in Section~\ref{sec:hp}, for $m_{\rm DM} < m_h /2$ collider searches for invisible Higgs decays are potentially competitive with direct detection constraints, and in Figure~\ref{fig:nonthermal} we plot the current bounds derived from limits on invisible Higgs decays at the LHC \cite{Khachatryan:2016whc}. Such searches are particularly significant for Higgs portal models with a non-thermal cosmological history, since they are more constraining than direct detection limits in some parts of the parameter space that lead to the required DM relic abundance. Additionally, there is a substantial region of viable parameter space close to the current bound. As a result, future improvement in these searches could lead to such a model being discovered.

The sensitivity of indirect searches for Higgs invisible decays at the LHC (from the measurement of the Higgs decay rates to SM particles)  will increase slightly  as more data is accumulated, and these could reach invisible branching fractions of order $\Gamma_{\rm invis}/\left(\Gamma_{\rm SM}+ \Gamma_{\rm invis} \right) \sim 0.1$. However, the potential improvement is limited by the large uncertainties in the cross section for Higgs production via gluon fusion due to QCD corrections \cite{Englert:2011aa,Dittmaier:2011ti,Djouadi:2011aa}. In contrast, the sensitivity of direct searches for invisible Higgs decays could improve dramatically over the course of the LHC's high luminosity run, and it is expected that invisible branching fractions as small as $\sim 0.02$ could be reached \cite{Eboli:2000ze,Bernaciak:2014pna,Biekotter:2017gyu}. Searches at a 100 TeV hadron collider could be even more sensitive, potentially down to invisible branching fractions of $\sim 0.005$ \cite{Goncalves:2017gzy,Dutta:2017sod}, although 
the unknown systematic uncertainties at such a machine makes it difficult to reliably predict the achievable sensitivity \cite{Goncalves:2017gzy}.\footnote{A Large Hadron-electron collider could reach invisible branching fractions of $\sim 0.07$ \cite{Tang:2015uha}.} 


A future $e^+ e^-$ collider with a centre of mass energy of $500~\GeV$ would also be extremely sensitive to invisible Higgs decays, and could detect an invisible branching fraction of $\sim 0.005$. In this case, the most powerful search channel is Higgs production in association with a Z-boson followed by invisible decay. Further, a study of the distribution of the missing invariant mass could confirm the Higgs portal nature of an observed signal \cite{AguilarSaavedra:2001rg,Djouadi:2007ik,Chacko:2013lna}, and interference effects might even allow a scalar DM Higgs portal to be distinguished from models of fermion DM interacting with the SM Higgs through a light mediator  \cite{Ko:2016xwd,Kamon:2017yfx}.

In Figure~\ref{fig:nonthermal} we show the part of parameter space that would be explored by collider searches that are sensitive to an invisible Higgs branching fraction of $1\%$, which is plausible at a future machine. For models with a relatively low reheating temperature $\sim 20~\MeV$, a significant proportion of the parameter space will be covered. This includes many models for which the DM relic abundance is set by production from the thermal bath without chemical equilibrium. Additionally, part of the parameter space in which DM is produced directly from modulus decays will also be reached in scenarios with relatively large modulus to DM branching fraction, for example $B \sim 10^{-3}$.

Conversely, collider searches for models with $m_{\rm DM} >  m_h/2$ are expected to remain weak. An $e^+ e^-$ collider could probe Higgs portal couplings of order $1$, which is far weaker than existing direct detection bounds in the minimal theories that we consider \cite{Chacko:2013lna}. Such searches could however be important in more complex models \cite{Kanemura:2010sh,Kamenik:2012hn,Craig:2014lda}, and they could also be significant in models of fermion Higgs portal DM if the additional states added to UV complete the portal operator are relatively light \cite{Freitas:2015hsa,Fedderke:2015txa,Ghorbani:2016edw,Westhoff:2016rcb}.

\subsection{Direct and indirect detection}

The sensitivity of direct detection experiments is also expected to increase substantially in the near future. However, such searches will eventually reach DM-nucleon cross sections at which the irreducible neutrino background, arising from coherent scattering of solar, atmospheric, and supernovae neutrinos, becomes problematic \cite{Billard:2013qya}. Improving the sensitivity of direct detection searches beyond this level is extremely challenging, and would require techniques such as directional detection \cite{Grothaus:2014hja}. 
In Figure~\ref{fig:nonthermal} we plot the neutrino floor, as well as the current direct detection constraints.

Relevant to the high DM mass region, the LZ experiment is expected to begin operation in 2020 \cite{Akerib:2015cja}. This is a next generation xenon experiment, which has an anticipated sensitivity that is close to the neutrino floor for DM masses $\gtrsim 6~\GeV$, after 1000 days of exposure assuming a $5.6$ tonne fiducial mass (the reach of the XENONnT experiment and the PandaX upgrade are expected to be similar). As well as exploring all of the resonance region in the thermal cosmological history case, this will cover a significant amount of the parameter space for which the DM relic abundance is set by freeze out during matter domination in non-thermal models with a relatively high reheating temperature $\sim 10~\GeV$.

At smaller DM masses SuperCDMS SNOLAB, which is based on cryogenic detectors, could reach DM nucleon scattering cross sections a factor of $\simeq 10$ (i.e. a factor of $\sim {\rm few}$ in the portal coupling) above the neutrino floor for DM masses in the range $0.7~\GeV \lesssim m_{\rm DM} \lesssim 10~\GeV$ \cite{Agnese:2016cpb}. From Figure~\ref{fig:nonthermal} left, such an experiment will explore an interesting part of the parameter space for non-thermal models that have a relatively low reheating temperature. The reach is comparable to that of future collider searches for Higgs invisible decays in this DM mass range, and, as mentioned, an observation in both of these channels would greatly increase the potential for discriminating between different DM models.\footnote{For $10~\GeV \lesssim m_{\rm DM} \lesssim m_h/2$ existing direct detection bounds already exclude most of the parameter space that could be probed by collider measurements in the foreseeable future.}

The prospects for improvement in the reach of indirect detection experiments are comparatively less dramatic, at least in the short term. The upcoming CTA experiment \cite{Silverwood:2014yza,Beniwal:2015sdl} might explore some parts of parameter space at high DM masses that are not currently excluded by direct detection experiments, and the potential reach is shown in Figure~\ref{fig:thermal}. However, as discussed in Section~\ref{sec:searches}, this is obtained assuming a relatively optimistic DM spatial distribution. Future measurements of anti-protons or anti-deuterium \cite{Donato:1999gy} produced by DM annihilations might also be sensitive to parts of parameter space that are not currently excluded \cite{Goudelis:2009zz,Urbano:2014hda}, however  astrophysical backgrounds and uncertainties are likely to remain significant.

\subsection{Astrophysical effects}

DM self-interactions  can be astrophysically relevant if the scattering cross section is of order $\sigma_s/m_{\rm DM} \sim \cm^2/{\rm g}$, and such interactions have been proposed as a possible resolution of apparent disagreements between numerical simulations of cold DM and observations of small-scale structure \cite{Spergel:1999mh}. Present discrepancies include the core vs cusp problem \cite{Navarro:1996gj,2011ApJ...742...20W}, the missing satellites problem \cite{2012MNRAS.422.1203B}, and the too-big-to-fail problem \cite{BoylanKolchin:2011de}, although it is plausible that baryonic effects such as photoionisation and supernova feedback could explain some or all of these \cite{2012MNRAS.422.1231G}.

Self interactions can be included in the scalar Higgs portal DM model simply by introducing the renormalisible interaction
\beq \label{eq:dmsi}
\mathcal{L} \supset -\lambda_s \chi^4 ~.
\eeq
This leads to $2 \rightarrow 2$ DM scattering with a rate that is unconnected to the dynamics that sets the DM relic abundance, and which can therefore be large \cite{Burgess:2000yq}. In the limit that $\lambda_s \gg \lambda$, which is the case in all of the viable parameter space with astrophysically relevant self-interactions, the DM scattering cross section is
\beq \label{eq:si1}
\frac{\sigma_s}{m_{\rm DM}} = \frac{9 \lambda_s^2}{2\pi m_{\rm DM}^3}= 1.4 \cm^2/{\rm g} \left(\frac{\lambda_s}{2\pi/3}\right)^2 \left(\frac{0.1 \GeV}{m_{\rm DM}} \right)^3~.
\eeq
Here $\lambda_s$ is normalised to its maximum allowed value such that the zeroth partial wave amplitude $a_0$ has $\left|{\rm Re}~ a_0 \right| < \frac{1}{2}$, as is required for perturbative unitarity \cite{Campbell:2015fra}. Consequently, astrophysically relevant DM self interactions are possible only for relatively small DM masses. This part of parameter space is robustly excluded in minimal models with a thermal cosmological history, however a non-thermal cosmology with a relatively low reheating temperature leads to viable models, for example as shown in Figure~\ref{fig:nonthermal} left.\footnote{Models with more complex hidden sectors, with the DM coupled to a light mediator that interacts with the visible sector through a Higgs portal operator, can lead to significant DM self-interactions. However, even in this case it is hard to evade observational constraints \cite{Feng:2009hw}. The mediator is required to decay before BBN, which rules out many models \cite{Kaplinghat:2013yxa}, although this is possible if it is coupled to relatively light right handed neutrinos \cite{Kouvaris:2014uoa}. CMB, direct detection, and invisible Higgs decay bounds can also be important.}

The DM self interactions from Eq.~\eqref{eq:dmsi} are short range contact interactions. These are potentially less appealing than the velocity dependent interactions generated by a light mediator \cite{Loeb:2010gj}, since there are relatively stringent constraints from galaxy clusters and halo ellipticity $\sigma_s/m_{\rm DM} \lesssim 1 \cm^2/{\rm g}$
 \cite{Markevitch:2003at,Randall:2007ph,Rocha:2012jg,Peter:2012jh,Harvey:2015hha}, which are ameliorated in light mediator models (due to the larger typical DM velocity in such systems). Despite this, short range self interactions could still play a role in resolving at least some of the discrepancies between simulations and observations, and they could also explain the separation between the DM halo and stars in a galaxy falling into the Abel 3827 galaxy cluster \cite{Kahlhoefer:2015vua}. Their effects might also be observationally distinguishable from those of the long range interactions generated by a light mediator.\footnote{DM self interactions could also lead to new indirect detection signals, for example neutrinos produced by DM annihilations in the sun \cite{Burgess:2000yq,Albuquerque:2013xna,Chen:2014oaa,Chen:2015bwa}. A preliminary analysis indicates that these are not relevant in phenomenologically viable parts of parameter space in the present model.}

As mentioned, if DM is dominantly produced from the visible sector thermal bath its temperature will be comparable to that of the visible sector. However, in parts of parameter space for which it is produced directly from modulus decays, the DM might never reach kinetic equilibrium with the visible sector if the portal coupling is extremely small \cite{Hisano:2000dz,Lin:2000qq,Gelmini:2006vn} (we leave a computation of the upper value of $\lambda$ for which thermalisation will not occur to future work). Each DM state is produced with a typical kinetic energy of $m_{\phi}/2$ (assuming that $m_{\phi} \gg m_{\rm DM}$), so in this regime the DM temperature is larger than that of the visible sector by a factor of $\sim m_{\phi} / T_{\rm RH}$. This could lead to observable effects on structure formation or violate constraints on warm DM, and Lyman-$\alpha$ observations  \cite{Viel:2005qj,Boyarsky:2008xj,Viel:2013apy,Boyarsky:2008ju,Horiuchi:2013noa} require that such models satisfy
\beq
\begin{aligned}
& \frac{T_t}{T_{\rm RH}} \frac{m_{\phi}}{m_{\rm DM}}  \lesssim 3 \times 10^{-8} ~,
\end{aligned}
\eeq
where $T_t$ is the temperature of the universe today (observations of dwarf spheroidal galaxies also lead to similar limits). Given the relation between the reheating temperature and the modulus mass, Eq.~\eqref{eq:phidecay}, this is equivalent to requiring that
\beq  \label{eq:warm}
\left( \frac{m_{\rm DM}}{\GeV} \right) \left( \frac{T_{\rm RH}}{\GeV} \right)^{1/3} \gtrsim 10 ~.
\eeq
Comparing with the parameter space plotted in Figure~\ref{fig:nonthermal}, Eq.~\eqref{eq:warm} bound is borderline dangerous for some theories with a low reheating temperature, $\sim 20~\MeV$, and a relatively large modulus to DM branching ratio.\footnote{If the DM does not even thermalise among itself these constraints could change significantly  \cite{Murgia:2017lwo}.} More optimistically, such models could lead to effects at a level that could be observed in the near future. However, a careful computation is likely to reveal that there are large parts of viable parameter space in the direct production regime for which the portal coupling is large enough that the DM thermalises with the visible sector. Eq.~\eqref{eq:warm} is not significant in models with a larger reheating temperature, since direct production leads to the required relic abundance for much larger DM masses in these. 

Conversely, in other dynamical regimes a non-thermal cosmological history can result in DM that is significantly colder than the visible sector \cite{Gelmini:2008sh}. This could happen if the DM relic abundance is set by production from the visible sector thermal bath, and the temperature at which the DM kinetically decouples from the visible sector is larger than the visible sector reheating temperature, so that reheating increases the visible sector temperature relative to that of the DM. If this occurs, the smallest objects formed during hierarchical structure formation might be much smaller than those obtained assuming a thermal cosmological history, which could potentially affect indirect detection signals. However, we leave a full calculation of the kinetic decoupling temperature, necessary to determine if such a scenario is realised in Higgs portal models, for future work.

\section{Conclusions} \label{sec:con}

In this paper we have studied the viable parameter space of Higgs portal DM in theories with a period of late time matter domination due to a light, gravitationally coupled, scalar. Such a modification of the universe's cosmological history is well motivated from string theory UV completions of the SM. These generically contain many moduli fields, which automatically lead to a period of matter domination if the scale of supersymmetry breaking is relatively low. In such models large regions of parameter space remain viable after imposing the DM relic density constraint, and direct detection and collider bounds, in contrast to if a thermal cosmological history is assumed. In such models the DM abundance can be dominantly set by production from the visible sector thermal bath with or without chemical equilibrium, or by direct production from modulus decays.

Higgs portal DM can lead to a wide range of signatures at direct, indirect, and collider experiments. In models with a fairly low reheating temperature, relatively light DM with $m_{\rm DM} \sim \GeV$ is possible, which can lead to astrophysically significant DM self-interactions. Additionally, models with DM masses $\lesssim 60~\GeV$ could lead to signals in both future collider searches for invisible Higgs decays and direct detection experiments. Meanwhile, if the reheating temperature is larger, heavy DM is possible without large couplings to the visible sector or additional new hidden sector states.

Although we have focused on the scalar Higgs portal DM model due to its interesting phenomenology, a similar increase in the allowed parameter space is likely to occur in other DM models, for example DM interacting with the visible sector through the Z portal. Additionally, a non-thermal cosmological history could also lead to new experimental signatures in fermion Higgs portal DM models with pseudoscalar couplings to the Higgs. Unlike scalar Higgs portal DM, direct detection bounds are extremely weak in this case, and dynamics that enhance the DM relic abundance compared to in a thermal cosmological history are possible in viable theories. These could allow for larger values of the portal coupling than in thermal models, leading to combinations of signals in collider, direct detection, and indirect detection experiments that would otherwise not be possible.

\section*{Acknowledgements}
I am grateful to Bobby Acharya and Martin Gorbahn for very useful discussions.

\section*{Note added}
In the final stages of preparing this manuscript, the
paper~\cite{Bernal:2018ins}, which also studies Higgs portal DM in models with a non-thermal cosmological history, was posted on the arXiv.

\appendix

\section{Evolution of the DM number density during matter domination} \label{sec:ap}
For completeness we briefly summarise the equations governing the evolution of the modulus and radiation bath energy densities, and the DM number density, during a non-thermal cosmological history. These are given in terms of the variables defined in Eq.~\eqref{eq:cmdef}.

Following \cite{Drees:2017iod} we introduce a dimensionless Hubble parameter $\tilde{H}$
\beq
\tilde{H} = \sqrt{\Phi + \frac{R}{A} + E_{X} \frac{X}{T_{\rm RH}}} ~ 
\eeq
where $E_{X} \simeq \sqrt{m_{\rm DM}^2 + 3 T^2 } $ is the average energy of a DM particle. It is also convenient to define a parameter related to the average energy that goes into DM mass for each modulus quanta that decays by
\beq
\tilde{B} = \frac{E_X}{m_{\phi}} B~ .
\eeq
The evolution of the modulus energy density and DM number density is given by
\beq
\begin{aligned}
 \tilde{H} \frac{d \Phi}{dA} &= - \sqrt{\frac{\pi^2 g\left(T_{\rm RH}\right)}{30}} A^{1/2} \Phi ~,\\
  \tilde{H} \frac{d X}{dA} &=  \sqrt{\frac{\pi^2 g\left(T_{\rm RH}\right)}{30}} \frac{T_{\rm RH} B}{m_{\phi}} A^{1/2} \Phi + 3 M_{\rm Pl} T_{\rm RH} A^{-5/2} \left<\sigma v \right> \left(X_{\rm eq}^2 - X^2 \right) ~,\\
\end{aligned}
\eeq
where $X_{\rm eq}= \frac{A^3}{T_{\rm RH}^2} n_{\rm eq}$, with $n_{\rm eq}$ the equilibrium DM number density at a given temperature. The thermally averaged annihilation cross section $\left<\sigma v\right>$ has a strong temperature dependence in the Higgs portal model, and is evaluated at the temperature of the radiation bath. 
The evolution of the temperature is given by
\begin{equation}
\begin{aligned}
  \frac{dT}{dA} = & - \frac{1}{1+ \frac{T}{3 g_S} \frac{d g_S}{dT}} \frac{T}{A}   \\
    &  + \frac{1}{\left( 1+ \frac{T}{3 g_S} \frac{d g_S}{dT} \right)} \frac{15 T_{\rm RH}^6}{2 \pi^2 g_S  M_{\rm Pl} H T^3 } \left(\pi \sqrt{\frac{g\left(T_{\rm RH}\right)}{90}} (1-\tilde{B}) \frac{\Phi}{A^4} + \frac{2 M_{\rm Pl} E_X \left<\sigma v\right>}{A^{7}} \left(X^2 - X_{\rm eq}^2 \right) \right)
\end{aligned}
  \end{equation}
where $g_S$ is the effective number of relativistic degrees of freedom in the visible sector that contribute to the entropy density, and the energy density in radiation $\rho_R = \frac{\pi^2}{30} g T^4$.

\bibliographystyle{JHEP}
\bibliography{reference}

\end{document}